\documentclass[twocolumn,floats,floatfix,aps,pra]{revtex4-2}

\usepackage{amsfonts,amssymb,amsmath}
\usepackage{physics}
\usepackage{color,calc}
\usepackage{bm}
\usepackage{amsbsy,amsmath}
\usepackage{array}
\usepackage{booktabs}
\usepackage{graphicx}
\usepackage{graphics}
\usepackage{eepic,epsfig}
\usepackage[table, dvipsnames]{xcolor}
\usepackage{colortbl}
\usepackage{subcaption}
\usepackage[utf8]{inputenc}
\usepackage[T1]{fontenc}

\usepackage[breaklinks,hidelinks,colorlinks=true,linkcolor=blue,citecolor=blue,filecolor=blue,urlcolor=blue]{hyperref}

\usepackage[dvipsnames]{xcolor}

\newcommand{\comment}[1]{}

\def\be{ \begin{equation} }
\def\ee{ \end{equation} }
\def\bea{ \begin{eqnarray} }
\def\eea{ \end{eqnarray} }
\def\bse{ \begin{subequations} }
\def\ese{ \end{subequations} }
\def\ba{ \begin{array} }
\def\ea{ \end{array} }
\def\bt{ \begin{tabular} }
\def\et{ \end{tabular} }

\long\def\/*#1*/{}

\begin{document}

\title{Dielectric permittivity of electron plasma with laser beat waves}

\author{Anahit H. Shamamian\textsuperscript{\hyperref[1]{1},\hyperref[2]{2},\hyperref[3]{3}}}
\author{Lekdar A. Gevorgian\textsuperscript{\hyperref[1]{1}}}
\email{ledkar@yerphi.am}
\author{Hayk L. Gevorgyan\textsuperscript{\hyperref[1]{1},\hyperref[4]{4},\hyperref[5]{5},\hyperref[6]{6}}}
\email{hayk.gevorgyan@aanl.am}
\affiliation{
\phantomsection\label{1}{\textsuperscript{1}Matinyan Center for Theoretical Physics, Alikhanyan National Laboratory (Yerevan Physics Institute), 2 Alikhanian Brothers St., 0036 Yerevan, Armenia}\\
\phantomsection\label{2}{\textsuperscript{2}Military Academy after Vazgen Sargsyan MoD RA, 155 Davit Bek St., 0090 Yerevan, Armenia}\\
\phantomsection\label{3}{\textsuperscript{3}Plekhanov Russian University of Economics, Yerevan Branch, 5/2 Arzumanyan St., 0038, Yerevan, Armenia}\\
\phantomsection\label{4}{\textsuperscript{4}Division for Quantum Technologies, Alikhanyan National Laboratory (Yerevan Physics Institute), 2 Alikhanian Brothers St., 0036 Yerevan, Armenia}\\
\phantomsection\label{5}{\textsuperscript{5}Experimental Physics Division, Alikhanyan National Laboratory (Yerevan Physics Institute), 2 Alikhanian Brothers St., 0036 Yerevan, Armenia}\\
\phantomsection\label{6}{\textsuperscript{6}Center for Quantum Technologies, Faculty of Physics, St. Kliment Ohridski University of Sofia, 5 James Bourchier Blvd., 1164 Sofia, Bulgaria}}


\date{\today }

\begin{abstract}
Within the framework of the kinetic method, the process of interaction of a rarefied electron plasma with laser beat waves is investigated taking into account the initial spread in the velocities of plasma particles. Using 
the expression for the spectral distribution function of plasma particles in the presence of laser beat waves, an anisotropic tensor of the plasma dielectric permittivity is obtained. In this case, an expression for the 
nonequilibrium current density induced by the perturbed field and the constitutive equation connecting the conductivity tensor with the dielectric permittivity are determined. Assuming that in equilibrium the particle 
distribution function has the form of the Maxwellian distribution function, the dielectric tensor is diagonalized. Analytical expressions are obtained for the longitudinal and transverse components of the dielectric permittivity.
\end{abstract}

\maketitle

\section{Introduction}

A theoretical study of a plasma interacting with two laser beams of the equal intensity, the frequencies of which are much higher than, and their difference is of the order of the plasma frequency, was carried out in the article \cite{tajima1979}. In the articles \cite{channell1982} devoted to the study of various processes occurring in plasma, these beams are called laser beat waves (LBWs). Basically, it is proposed to use such intense laser beams to accelerate particles in plasma.

Plasma studies in the presence of LBWs were carried out in the hydrodynamic approximation \cite{ruth1982}. The interest in further investigation of this process is associated with the fact that the interaction of plasma with LBWs leads to a nonlinear change in the plasma density with a linearly varying amplitude 
\cite{ruth1982, gevorgian1991}.

In \cite{gevorgian1996}, a kinetic description of the process of interaction of a plasma with LBWs is presented. Here the expressions for spectral distribution function and spectral density are obtained. Note that the process of radiation of a high-current electron beam in the presence of LBWs was further investigated. 
Coherent effects in relativistic electron beams radiation \cite{HGevorgyan2017, HGevorgyan2021a, HGevorgyan2021b, HGevorgyan2015} in the presence of LBWs are investigated in \cite{gevorgian1993, shamamian2007, shamamian2005, shamamian2010}. Generation of THz radiation from beating of two intense laser beams in magnetized plasma are considered in \cite{bhasin2016, purohit2019}.

This article is devoted to further investigation of the process of interaction of plasma with LBWs. The study was carried out within the framework of the kinetic method for describing plasma, taking into account the initial spread in the velocities of plasma particles. Using the material equation for the conductivity tensor and the permittivity tensor \cite{alexandrov1978, silin1961}, the anisotropic dielectric tensor of the plasma with LBWs is determined.

Further investigation was carried out under the assumption that in the absence of LBWs, the plasma has a Maxwellian distribution of particles. Using the well-known expressions of Fok's article \cite{fok1946}, the anisotropic dielectric tensor is diagonalized when the waves excited in the plasma propagate in the longitudinal direction along the LBWs’ motion.

\section{Plasma - Laser Beat Waves interaction}
Let us consider the process of interaction of collisionless nonrelativistic plasma with LBWs propagating along the $\text{OZ}$ axis.

LBWs are the result of the interference of two laser beams of the equal intensity, the frequencies of which $\omega_1$ and $\omega_2$ are approximately equal and much higher than the plasma frequency $\omega_p$ \cite{tajima1979, channell1982, ruth1982, gevorgian1991, gevorgian1996}:
\be\label{}
\begin{split}
    \omega_1, \omega_2 &\approx \omega_L,\\
    \omega_1, \omega_2 &\gg \omega_p.
\end{split}
\ee

In this case, the difference between the frequencies of the laser beams is approximately equal to the plasma frequency
\be\label{}
\omega_1 -\omega_2 \approx \omega_p.
\ee

LBWs entering the plasma generate plasma waves with a plasma frequency $\omega_p$ and a wave number $k_p$
\be\label{}
k_1 - k_2 \approx k_p .
\ee
where $k_1$ and $k_2$ wave vector lengths of laser beams.

In the hydrodynamic approximation, the interaction of plasma with LBWs leads to the following nonlinear change in the plasma density with a linearly varying amplitude [3-4]
\be\label{plasma-density}
\begin{split}
n(z, t) &= \alpha (k_p z - \omega_p t) \sin{(k_p z - \omega_p t)}, \\ 
\alpha &= - n_0 \frac{e^2 E^2_L}{4 m \omega_L^2 v^2_{ph}}.
\end{split}
\ee
where $n_0$ is the initial plasma density, $v_{ph} = \omega_p/k_p$ is the phase velocity of plasma waves with a plasma frequency $\omega_p = \left(4\pi n_0 e^2 \right)/m$, $E_L$ is the electric field strength of the laser beam, $e$ and $m$ are charge and mass of an electron.

Using the collisionless kinetic equation of Boltzmann-Vlasov, the following expression for the spectral distribution function of a rarefied nonrelativistic plasma in the presence of LBWs was determined \cite{gevorgian1996}
\be\label{spectral-dist-function}
\begin{split}
    f = f(\vb*{p}, \vb*{k}, \omega) & = f_0 + A_L \frac{\omega_p \vb*{z}}{k_p v_z - \omega_p} \frac{\partial f_0}{\partial \vb*{p}}, \\
    A_L &= \frac{e^2 E^2_L}{4 m \omega^2_L v_{ph}},
\end{split}
\ee
where the value $A_L$ is defined as the parameter of LBWs, $v_z$ is the $\text{z}$ component of the velocity of plasma, which is moving, like the LBWs, along the $\text{OZ}$ axis. Note that a possible transition between  expressions \eqref{plasma-density} and \eqref{spectral-dist-function} is shown in \cite{gevorgian1996}.

\section{Dielectric permittivity of plasma in the presence of LBWs}
The initial state of the plasma in the presence of LBWs is described by the spectral distribution function \eqref{spectral-dist-function} which is determined in \cite{gevorgian1996}. We use this expression to find the permittivity tensor. 

Within the framework of perturbation theory, we obtain an expression for the perturbation of the distribution function of the initial state of the plasma in the presence of LBWs due to the action of small perturbation $\vb*{E}_{per}$ electric and $\vb*{B}_{per}$ magnetic fields, the appearance of which is due to fluctuation perturbation of the initial state of the plasma \cite{alexandrov1978, silin1961}
\be\label{pert-dist-function}
F = f + \delta f,
\ee
where $\delta f$ is the perturbation of the distribution function due to the fields $\vb*{E}_{per}$ and $\vb*{B}_{per}$. Note that, we will assume the perturbation of the distribution function, as well as the values of the perturbed fields of the same order of smallness.

The collisionless kinetic equation of Boltzmann-Vlasov for the perturbed plasma distribution function in the presence of LBWs and perturbed fields has the form
\be\label{plasma-dist-function-LBW}
\begin{split}
    \frac{\partial F}{\partial t} &+ \vb*{v} \frac{\partial F}{\partial \vb*{r}} + e \left(\vb*{E}_L + \frac{[\vb*{v} \times \vb*{B}_L]}{c} \right) \frac{\partial F}{\partial \vb*{p}} + \\
    & + e \left(\vb*{E}_{per} + \frac{[\vb*{v} \times \vb*{B}_{per}]}{c} \right) \frac{\partial F}{\partial \vb*{p}},
\end{split}
\ee
where $\vb*{E}_L$ and $\vb*{B}_L$ are the vectors of the strengths of the electric and magnetic fields of the laser beams, $\vb*{v}$ is the nonrelativistic velocity of plasma particles moving in the same direction with the LBWs.

Substituting expression \eqref{pert-dist-function} into \eqref{plasma-dist-function-LBW} and taking into account the fact that the laser fields are transversely polarized, as well as the fact that the derivative of the distribution function entering into the third and fourth terms of equation \eqref{plasma-dist-function-LBW} is determined by the direction of the momentum, one can obtain the following equation for the perturbation of the distribution function
\be\label{}
\frac{\partial \delta f}{\partial t} + \vb*{v} \frac{\partial \delta f}{\partial \vb*{r}} + e \vb*{E}_{per} \frac{\partial f}{\partial \vb*{p}} = 0,
\ee
where terms of the second and higher orders of smallness are not taken into account.

We represent the dependence on time and coordinates of the perturbed quantities included in the equation in the form $e^{i (\vb*{k} \cdot \vb*{r} - \omega t)}$, then we obtain
\be\label{}
\delta f = i e \frac{\vb*{E}_{per}}{\vb*{k}\cdot \vb*{v} - \omega} \frac{\partial f}{\partial \vb*{p}},
\ee
where $\omega$ and $\vb*{k}$ are the frequency and wave vector of plane waves propagating in the plasma.

The nonequilibrium current density, induced by the perturbed field \cite{alexandrov1978},
\be\label{}
\delta j_i = e \int v_i \delta f d \vb*{p} = i e^2 \int v_i \frac{\vb*{E}_{per}}{\vb*{k} \cdot \vb*{v} - \omega} \frac{\partial f}{\partial \vb*{p}} d \vb*{p},
\ee
according to the material equation:
\be\label{}
\delta j_i = \sigma_{ij}{\left(\vb*{E}_{per}\right)_j}, 
\ee
determines conductivity tensor $\sigma_{ij}$ as follows
\be\label{conduct-tensor}
\sigma_{ij} = \sigma_{ij} (\vb*{k}, \omega) = i e^2 \int \frac{v_i}{\vb*{k} \cdot \vb*{v} - \omega} \left(\frac{\partial f}{\partial \vb*{p}}\right)_j d\vb*{p} ,
\ee
where $\delta j_i$ and $v_i$, respectively, are the i-th components of the vectors of current density $\delta \vb*{j}$ and velocity $\vb*{v}$, respectively.

We use the following equation connecting the complex conductivity and the complex permittivity
\be\label{complex-permitt}
\varepsilon_{ij} (\vb*{k}, \omega) = \delta_{ij} (\vb*{k}, \omega) + \frac{4 \pi i}{\omega} \sigma_{ij} (\vb*{k}, \omega),
\ee
where $\delta_{ij} = \delta_{ij} (\vb*{k}, \omega)$ is the Kronecker symbol.

Substituting expression \eqref{conduct-tensor} into connection equation \eqref{complex-permitt}, we obtain the following expression for the anisotropic dielectric tensor of plasma in the presence of LBWs
\be\label{dielectric-tensor-LBW}
\varepsilon_{ij} (\vb*{k}, \omega) = \delta_{ij} (\vb*{k}, \omega) - \frac{4\pi e^2}{\omega} \int \frac{v_i}{\vb*{k}\cdot \vb*{v} - \omega} \frac{\partial f}{\partial \vb*{p}} d \vb*{p}.
\ee

Since the initial state of the plasma interacting with LBWs is described by the spectral distribution function \eqref{plasma-dist-function-LBW}, therefore, substituting this expression into \eqref{dielectric-tensor-LBW}, we obtain the following expression for the anisotropic dielectric permittivity tensor:
\be\label{anisotrop-diel-permit}
\begin{split}
\varepsilon_{ij} (\vb*{k}, \omega) &= \delta_{ij} (\vb*{k}, \omega) - \\
& -\frac{4\pi e^2}{\omega} \int \frac{v_i}{\vb*{k}\cdot \vb*{v} - \omega} \frac{\partial}{\partial \vb*{p}} \left(f_0 - A_L \frac{\partial f_0}{\partial p_z} \right) d \vb*{p}.    
\end{split}
\ee

\section{Maxwellian distribution of plasma particles: stationary plasma}
Suppose that in the absence of LBWs, the particle distribution is represented by the Maxwell distribution function
\be\label{Maxwell-dist-func}
f_0 = f_0 (\vb*{p}) = \frac{n_0}{(2\pi m k T)^{3/2}} e^{-\frac{p^2_x}{2 m k T}} e^{-\frac{p^2_y}{2 m k T}} e^{-\frac{p^2_z}{2 m k T}},
\ee
where $T$ is the plasma temperature, $k$ is the Boltzmann constant, $p_x$, $p_y$, $p_z$ are the components of the momentum vector $\vb*{p}$ in the direction of the coordinate axes of the Cartesian coordinate system.

In this case, the tensor of the dielectric permittivity of a stationary plasma interacting with LBWs \eqref{anisotrop-diel-permit} can be diagonalized if we consider the longitudinal propagation of waves excited in the plasma ($\vb*{k}_x = \vb*{k}_y = 0$).

Therefore, the longitudinal $\varepsilon_l (\vb*{k}, \omega)$ and transverse $\varepsilon_{tr} (\vb*{k}, \omega)$ components of the dielectric 
permittivity have the following form
\be\label{long-diel-permit}
\begin{split}
&\varepsilon_l(\vb*{k}, \omega) = 1 - \frac{4\pi e^2}{\omega} \int \frac{v_z}{k_z v_z - \omega} \frac{\partial f}{\partial p_z} d\vb*{p} = \\
&= 1 - \frac{4\pi e^2}{\omega} \int \frac{v_z}{k_z v_z - \omega} \frac{\partial}{\partial p_z} \left(f_0 - A_L \frac{\partial f_0}{\partial p_z} \right) d\vb*{p},    
\end{split}
\ee
\be\label{trans-diel-permit}
\begin{split}
&\varepsilon_{tr}(\vb*{k}, \omega) = 1 - \frac{4\pi e^2}{\omega} \int \frac{v_z}{k_z v_z - \omega} \frac{\partial f}{\partial p_x} d\vb*{p} = \\
&= 1 - \frac{4\pi e^2}{\omega} \int \frac{v_z}{k_z v_z - \omega} \frac{\partial}{\partial p_x} \left(f_0 - A_L \frac{\partial f_0}{\partial p_x} \right) d\vb*{p}.    
\end{split}
\ee

Substituting expression \eqref{Maxwell-dist-func} into \eqref{long-diel-permit} and \eqref{trans-diel-permit}, after simple transformations we have the following expressions for the longitudinal and transverse components of the permittivity
\begin{widetext}
\be\label{long-diel-permit-1}
    \varepsilon_l (\vb*{k}, \omega) = 1 - \frac{\omega^2_p}{(2\pi)^{3/2} (m \chi T)^{5/2} \omega} \left[\int \left(A_L p_z - p^2_z - \frac{A_L p^3_z}{m \chi T} \right) \frac{e^{- \frac{p^2_z}{2 m k T}}}{k_z v_z - \omega} d p_z\right] \left[\int e^{-\frac{p^2_x}{2 m k T}} d p_x\right] \left[\int e^{-\frac{p^2_y}{2 m k T}} d p_y\right], \\
\ee
\be\label{trans-diel-permit-1}
    \varepsilon_{tr} (\vb*{k}, \omega) = 1 - \frac{\omega^2_p}{(2\pi)^{3/2} (m \chi T)^{5/2} \omega} \left[\int \left(1 + \frac{A_L p_z}{m \chi T}\right) \frac{e^{- \frac{p^2_z}{2 m k T}}}{k_z v_z - \omega} d p_z\right] \left[\int p^2_x e^{-\frac{p^2_x}{2 m k T}} d p_x\right] \left[\int e^{-\frac{p^2_y}{2 m k T}} d p_y\right]. 
\ee

After calculating the integrals taken in the infinite limits from $-\infty$ to $+\infty$, expressions \eqref{long-diel-permit-1} and \eqref{trans-diel-permit-1} are transformed as follows
\be\label{}
\varepsilon_l (\vb*{k}, \omega) = 1 - \frac{\omega^2_p}{(2\pi)^{1/2} (m \chi T)^{3/2} \omega} \left[\int \left(A_L p_z - p^2_z - \frac{A_L p^3_z}{m \chi T} \right) \frac{e^{- \frac{p^2_z}{2 m k T}}}{k_z v_z - \omega} d p_z\right], 
\ee
\be\label{}
\varepsilon_{tr} (\vb*{k}, \omega) = 1 - \frac{\omega^2_p}{(2 \pi m \chi T)^{1/2} \omega} \left[\int \left(1 + \frac{A_L p_z}{m \chi T}\right) \cdot \frac{e^{- \frac{p^2_z}{2 m k T}}}{k_z v_z - \omega} d p_z  \right], 
\ee

Using the following notation:
\be\label{}
   y = \frac{p_z}{\sqrt{m \chi T}}, \quad
   V_T = \sqrt{\frac{\chi T}{m}}, \quad
   \beta = \frac{\omega}{k c}, \quad
   \beta_T = \frac{V_T}{c}, \quad
   \frac{\beta}{\beta_T} = \frac{\omega}{k_z v_z},
\ee
and the well-known results of Fok's work \cite{fok1946}, for the longitudinal and transverse components, finally we have
\be\label{long-diel-permit-2}
    \varepsilon_l (\vb*{k}, \omega) = 1 - \frac{\omega^2_p \left(\frac{\beta}{\beta_T}\right)^2}{(2\pi)^{1/2} (m \chi T)^{3/2} \omega} \left[\left(1 - I \left(\frac{\beta}{\beta_T}\right)\right) \left(1 + \frac{A_L}{\sqrt{m\chi T}} \frac{\beta}{\beta_T}\right) + \frac{A_L}{\sqrt{m \chi T \frac{\beta}{\beta_T}}} I\left(\frac{\beta}{\beta_T}\right) \right], 
\ee
and 
\be\label{trans-diel-permit-2}
\varepsilon_{tr} (\vb*{k}, \omega) = 1 - \frac{\omega^2_p}{\omega^2} \left[\left(1 + \frac{A_L}{\sqrt{m\chi T}} \frac{\beta}{\beta_T} \right) I\left(\frac{\beta}{\beta_T}\right) - \frac{A_L}{\sqrt{m \chi T}} \frac{\beta}{\beta_T}\right].
\ee
In the last \eqref{long-diel-permit-2}, \eqref{trans-diel-permit-2} 
expressions $I \left(\frac{\beta}{\beta_T}\right)$ is the following function
\be\label{}
I\left(\frac{\beta}{\beta_T}\right) = \frac{\beta}{\beta_T} \frac{I_+ \left(\frac{\beta}{\beta_T}\right)}{\sqrt{2\pi}} = \frac{\beta}{\beta_T} e^{- \left(\frac{\beta}{\beta_T}\right)^2} \int\limits_{i \infty}^{\frac{\beta}{\beta_T}} e^{\frac{\tau^2}{2}} d\tau, \quad I_+ \left(\frac{\beta}{\beta_T}\right) = \sqrt{2\pi} e^{-\left(\frac{\beta}{\beta_T}\right)} \int\limits_{i\infty}^{\frac{\beta}{\beta_T}} e^{\frac{\tau^2}{2}} d\tau .
\ee
\end{widetext}

In the future, when studying the spectra of longitudinal and transverse waves propagating in plasma in the presence of LBWs, one can use the asymptotic values of the function $I \left(\frac{\beta}{\beta_T}\right)$ associated 
with the probability integral \cite{abramowitz1968}.

\section{Conclusion}
An expression is obtained for the anisotropic tensor of the dielectric permittivity of a plasma interacting with LBWs. Assuming the distribution of plasma particles to be Maxwellian in the absence of LBWs, the anisotropic tensor is diagonalized when the waves excited in the plasma propagate along 
the movement of LBWs. As a result, analytical expressions are obtained for the longitudinal and transverse components of the dielectric permittivity, allowing further research in asymptotic approximations in different frequency ranges.

In the case of an isotropic dielectric tensor for excited and propagating waves in a plasma, the known dispersion equation decomposes into two separate equations for both longitudinal and transverse waves. In the propagation of waves excited in the direction of LBWs in the plasma, due to the isotropy of the tensor, analogous equations can be obtained. This will make it possible to study the spectra of longitudinal and transverse waves excited in the plasma, which will differ significantly from the spectra in the absence of LBWs.

The results obtained for a stationary plasma can also be used to study a high-current electron bunch with density modulation induced by LBWs. Radiation of such bunches in different frequency ranges are of interest.

The main idea of the paper, \emph{the plasma interacting with LBWs}, can be considered in the case of  plasma's natural anisotropy like in the papers \cite{cremaschini2011, cremaschini2023a, cremaschini2023b, cremaschini2024}, and it will be interesting to study the state of this anisotropy in more detail.

The paper was first presented at the conference in 2021, organized in Yerevan and Meghri, Armenia \cite{shamamian2021}.







\begin{thebibliography}{99}


\bibitem{tajima1979} T. Tajima and J.M. Dawson, ``Laser electron accelerator,'' \href{https://doi.org/10.1103/physrevlett.43.267}{Phys. Rev. Lett. \textbf{43}, 267 (1979)}.

\bibitem{channell1982}
P.J. Channell, \emph{Laser acceleration of particles}, AIP Conf. Proc. \textbf{91}, CONF-820241- (Los Alamos, 1982).

\bibitem{ruth1982}
R.D. Ruth and A.W. Chao, ``Plasma laser accelerator: Longitudinal dynamics, the plasma/laser interaction, and a qualitative design,'' in P.J. Channell, \emph{Laser acceleration of particles}, AIP Conf. Proc. \textbf{91} (1), pp. 94--111 (1982).

\bibitem{gevorgian1991}
L.A. Gevorgian and A.H. Shamamian, ``Coherent effects in relativistic electron beams radiation in the presence of beat waves,'' Preprint YERPHI-\textbf{1321} (16)-91, Yerevan, 1--18 (1991).

\bibitem{gevorgian1996}
L.A. Gevorgian and A.H. Shamamian, ``Kinetic description of plasma interaction process with laser beat waves,'' Izv. NAN Armenia, Fizika \textbf{31} (2), 47--54 (1996).

\bibitem{HGevorgyan2017}
H.L. Gevorgyan and L.A. Gevorgian, 
``Coherent radiation characteristics of modulated electron bunch formed in stack of two plates,''
\href{https://doi.org/10.1016/j.nimb.2017.02.079}{Nucl. Instr. and Meth. B \textbf{402} (1), 126--129 (2017)}.

\bibitem{HGevorgyan2021a}
H.L. Gevorgyan, ``X-ray crystalline undulator radiation in water window,'' \href{https://doi.org/10.52853/18291171-2021.14.2-105}{Arm. J. Phys. \textbf{14}, 105--109 (2021)}.

\bibitem{HGevorgyan2021b}
H.L. Gevorgyan, L.A. Gevorgian, and A.H. Shamamian, ``Gain and features of radiation of positrons in a crystalline undulator with sections,'' \href{https://doi.org/10.3103/S1068337221030117}{J. Contemp. Phys. \textbf{56}, 159--168 (2021)}.

\bibitem{HGevorgyan2015}
K. Gevorgyan, L. Gevorgian, and H. Gevorgyan, ``Positron bunch radiation in the system of tightly-packed nanotubes,'' \href{https://doi.org/10.48550/arXiv.1512.08282}{arXiv:1512.08282 (2015)}.

\bibitem{gevorgian1993}
L.A. Gevorgian and A. Shamamian, ``Coherent amplification of radiation of electron beams interacting with laser beat waves,'' Int. J. Mod. Phys. A \textbf{28}, 1175 (1993).

\bibitem{shamamian2007}
A. Shamamian and L. Gevorgian, ``Coherent sub-millimeter undulator radiation from modulated electron beam,'' \href{https://doi.org/10.1117/12.742068}{Proc. SPIE \textbf{6634}, 66341C (2007)}.

\bibitem{shamamian2005}
A.H. Shamamian, ``Coherent radiation of the modulated electron bunch,'' \href{https://doi.org/10.1117/12.640018}{Proc. SPIE \textbf{5974}, 59740Y (2005)}.

\bibitem{shamamian2010}
A. Shamamian and L. Gevorgian, ``Hard photons powerful radiation of electron bunch interacting with plasma beat wayes,'' in S.B. Dabagov and L. Palumbo (Eds.), \href{https://doi.org/10.1142/9789814307017_0053}{\emph{Charged And Neutral Particles Channeling Phenomena: Channeling 2008}}, Proc. of the 51st Workshop of the INFN ELOISATRON Project, Erice, Italy, (World Scientific, Singapore, 2010), Vol. \textbf{30}, pp. 581--586.

\bibitem{bhasin2016}
L. Bhasin, V.K. Tripathi, and P. Kumar, ``Laser beat wave resonant terahertz generation in a magnetized plasma channel,'' \href{https://doi.org/10.1063/1.4940962}{Phys. Plasmas \textbf{23}, 023101 (2016)}. 

\bibitem{purohit2019}
G. Purohit, V. Rawat, and P. Rawat, ``Generation of terahertz radiation from beating of two intense cosh-Gaussian laser beams in magnetized plasma,'' \href{https://doi.org/10.1017/s0263034619000685}{Laser and Part. Beams \textbf{37}, 415 (2019)}.

\bibitem{alexandrov1978}
A.F. Alexandrov, A.S. Bogdankevich, and A.A. Rukhadze, \emph{Principles of Plasma Electrodynamics} (Vysshaya Shkola, Moscow, 1978) [in Russian]; Vol. \textbf{9} (Springer, Berlin, 1984).

\bibitem{silin1961}
V.P. Silin and A.A. Rukhadze, \emph{Electromagnetic Properties of Plasma and Plasma-like Media}, 
(Atomizdat, Moscow, 1961) [in Russian].

\bibitem{fok1946}
V.A. Fok, \emph{Diffraction of Radio Waves around the Surface of the Earth}, (Akad. Nauk SSSR, 
Moscow, 1946) [in Russian].
 
\bibitem{abramowitz1968}
M. Abramowitz and I.A. Stegun, \emph{Handbook of mathematical functions with formulas, graphs, and mathematical tables}, U.S. Department of Commerce, National Bureau of Standards Appl. Math. Series 55 (Vol. 55) (U.S. Government printing office, 1968).

\bibitem{cremaschini2011}
C. Cremaschini and M. Tessarotto, ``Kinetic description of rotating Tokamak plasmas with anisotropic temperatures in the collisionless regime,'' \href{https://doi.org/10.1063/1.3656978}{Phys. Plasmas \textbf{18}, 112502 (2011)}.

\bibitem{cremaschini2023a}
C. Cremaschini, J. Kovář, Z. Stuchl\'ik, and M. Tessarotto, ``Polytropic representation of the kinetic pressure tensor of non-ideal magnetized fluids in equilibrium toroidal structures,'' \href{
https://doi.org/10.1063/5.0134320}{Phys. Fluids \textbf{35}, 017123 (2023)}.

\bibitem{cremaschini2023b}
C. Cremaschini, ``Polytropic representation of non-isotropic kinetic pressure tensor for non-ideal plasma fluids in relativistic jets,'' \href{https://doi.org/10.1063/5.0154814}{Phys. Fluids \textbf{35}, 067101 (2023)}.

\bibitem{cremaschini2024}
C. Cremaschini and J. Kovář, ``Statistical characterization of the collective synchrotron radiation power emitted by non-ideal magnetized plasma fluids in relativistic jets,'' \href{https://doi.org/10.1063/5.0190676}{Phys. Fluids \textbf{36}, 037123 (2024)}.

\bibitem{shamamian2021}
A.H. Shamamian, L.A. Gevorgian, H.L. Gevorgyan, ``Dielectric permittivity of an electron plasma with laser beat waves,'' in \href{https://conference.iapp.am/publications/proceedings/}{\emph{Proceedings of International Conference on Electron, Positron, Neutron and X-ray Scattering under the External Influences}}, (Yerevan -- Meghri, Armenia, 2021), Part I, pp. 72--78.









\end{thebibliography}
\end{document}